\documentclass[12pt,showpacs]{revtex4}

\usepackage{amssymb}
\usepackage{amsmath}
\usepackage{graphicx}

\newcommand{\bq}{\begin{eqnarray}}
\newcommand{\eq}{\end{eqnarray}}
\newcommand{\bqn}{\begin{eqnarray*}}
\newcommand{\eqn}{\end{eqnarray*}}

\begin{document}
\title{Direct correlation functions of the Widom-Rowlinson model}

\author{R. Fantoni\footnote{Address: Dipartimento di Fisica Teorica 
dell' Universita` degli Studi di Trieste, Strada Costiera 11, 34014
Trieste, Italy; Phone: +39 040 2240608; Fax: +39 040 224601;
Electronic address: {\rm rfantoni@ts.infn.it}}}
\affiliation{Dipartimento di Fisica Teorica dell' Universit\`a  
and Istituto Nazionale di Fisica della Materia, Strada Costiera 11, 
34014 Trieste, Italy}
\author{G. Pastore\footnote{Address: Dipartimento di Fisica Teorica 
dell' Universita` degli Studi di Trieste, Strada Costiera 11, 34014
Trieste, Italy; Phone: +39 040 2240242; Fax: +39 040 224601; 
Electronic address: {\rm pastore@ts.infn.it}}}
\affiliation{Dipartimento di Fisica Teorica dell' Universit\`a  
and Istituto Nazionale di Fisica della Materia, Strada Costiera 11, 
34014 Trieste, Italy}
\date{\today}

\begin{abstract}
\noindent
We calculate, through Monte Carlo numerical simulations, the partial
total and direct correlation functions of the three dimensional
symmetric Widom-Rowlinson mixture. We find that the differences
between the partial direct correlation functions from simulation and
from the Percus-Yevick approximation (calculated analytically by Ahn
and Lebowitz) are well fitted by Gaussians. 
We provide an analytical expression for the fit parameters as function
of the density. We also present Monte Carlo simulation  data for the
direct correlation functions of a couple of non additive hard sphere
systems to discuss the modification induced by finite like diameters. 
\end{abstract}

\pacs{61.20.Ja,61.20.Gy}

\maketitle

{\bf KEYWORDS:} Widom-Rowlinson model, direct correlation function.
\section{Introduction}

Fluid binary mixtures may exhibit the phenomenon of phase separation.
The simplest system able to undergo a demixing phase transition is the 
model introduced by Widom and Rowlinson some years ago \cite{WR}.
Consider a binary mixture of non-additive hard spheres. This is a
fluid made of hard spheres of specie 1 of
diameter $R_{11}$ and number density $\rho_1$ and hard spheres of
specie 2 of diameter $R_{22}$ and number density $\rho_2$, with a pair
interaction potential between species $i$ and $j$ that can be written
as follows 
\bq \label{pp}
v_{ij}(r)=\left\{\begin{array}{ll}
\infty & r<R_{ij}\\
0      & r>R_{ij}
\end{array}\right.~~,
\eq
where $R_{12}=(R_{11}+R_{22})/2+\alpha$. The Widom-Rowlinson (WR) model is
obtained choosing the diameters of the spheres equal to 0,
\bq
R_{11}=R_{22}=0~~,
\eq
so that there is no interaction between  like spheres and there is a
hard core repulsion of diameter $\alpha$ between unlike spheres. The
symmetry of the system induces the symmetry of the unlike correlations
[$h_{12}(r)=h_{21}(r),~~ c_{12}(r)=c_{21}(r)$].
The WR model has been studied in the past by exact \cite{Ruelle} and
approximate \cite{Ahn73,Ahn74,Bergmann,KS79} methods  and it has been
shown that it exhibits a phase transition  at high density.  
More recently, additional studies have appeared and theoretical
predictions have been confirmed by Monte Carlo (MC) computer simulations
\cite{LHK90,DS95,Shew96,Johnson97} 

In this paper we will study the three dimensional symmetric Widom-Rowlinson
mixture for which $\rho_1=\rho_2=\rho/2$, where $\rho$ is the total
number density of the fluid, and
\bq
h_{11}(r)&=&h_{22}(r)~~,\\
c_{11}(r)&=&c_{22}(r)~~.
\eq
Moreover we know from (\ref{pp}) that the partial pair
correlation function $g_{ij}=h_{ij}+1$ must obey
\bq \label{g12rlalpha}
g_{ij}(r)=0~~~~\mbox{for $r<R_{ij}$}~~.
\eq

Our main goal is to focus on the direct correlation functions (dcf) of
the WR model as a simplified prototype of  non-additive hard spheres
(NAHS) systems. The reasons to focus on the dcf's is twofold: on the
one hand, they are easier functions to model and fit. On the other
hand, they play a central role in approximate theories like the
Percus-Yevick approximation or mean spherical approximation (MSA)
\cite{Hansen}. We hope that a better understanding of the dcf's
properties in the WR model, could  help in developing accurate analytical 
theories for the NAHS systems. 

We   calculate through Monte Carlo simulations the like
$g^{(MC)}_{11}(r)$ and unlike $g^{(MC)}_{12}(r)$ pair distribution
functions for a system large enough to allow a meaningful
determination of the correspondent partial direct
correlation functions $c^{(MC)}_{11}(r)$ and $c^{(MC)}_{12}(r)$, using
the Ornstein-Zernike equation \cite{Hansen}. We compare the results
for the unlike direct correlation function
with  the results of the Percus-Yevick (PY)
analytic solution found by Ahn and Lebowitz \cite{Ahn73,Ahn74}. In
the same spirit as the work of Grundke and Henderson for a mixture of
additive hard spheres \cite{Grundke72}, we propose a fit for the functions 
$\Delta^c_{11}(r)=c^{(MC)}_{11}(r)$ and
$\Delta^c_{12}(r)=c^{(MC)}_{12}(r)-c^{(PY)}_{12}(r)$. 

At the end of the paper we also show the results from two Monte Carlo
simulations on a mixture of non-additive hard spheres with equal
diameter spheres $R_{11}=R_{22}=R_{12}/2$ and on one with different
diameter spheres $R_{11}=0$ and $R_{22}=R_{12}$ to study the effect
of non zero like diameters on the WR dcf's. 
\section{Monte Carlo simulation and PY solution}

The Monte Carlo simulation was performed with a standard NVT
Metropolis algorithm \cite{AT} using $N=4000$ particles.  
Linked lists  \cite{AT} have been used to reduce the computational cost.
We generally used $5.2\times 10^8$ 
Monte Carlo steps where one step corresponds to the attempt to move a
single particle. The typical CPU time for each density was around 20
hours (runs at higher densities took longer than runs at smaller
densities) on a Compaq AlphaServer 4100 5/533.

We run the simulation of WR model at $6$ different densities
$\bar{\rho}=\rho\alpha^3=0.28748$, $0.4$, $0.45$, $0.5$, $0.575$, and
$0.65$. Notice that the most recent computer simulation calculations
\cite{Shew96,Johnson97}  
give consistent estimates of the critical density around 0.75. Our
data at the highest density (0.65) are consistent with a one phase
system. 

The Monte Carlo simulation returned the $g_{ij}(r)$ over a range not
less than $9.175\alpha$ for the densest system. In all the studied
cases the pair distribution functions attained their asymptotic value
well inside the maximum distance they were evaluated. Thus, it has
been possible to obtain accurate fourier transforms of the correlation
functions [$h_{ij}(k)$]. To obtain the $c_{ij}(r)$ we used
Ornstein-Zernike equation as follows 
\bq
c_{11}(k)&=&\frac{h_{11}(k)\left[1+\frac{\rho}{2}h_{11}(k)
\right]-\frac{\rho}{2}h_{12}^2(k)}
{\left[1+\frac{\rho}{2}h_{11}(k)\right]^2-
\left[\frac{\rho}{2}h_{12}(k)\right]^2}\\
c_{12}(k)&=&\frac{h_{12}(k)}{\left[1+\frac{\rho}{2}h_{11}(k)\right]^2
-\left[\frac{\rho}{2}h_{12}(k)\right]^2}
\eq
From the $h_{ij}(k)$ and $c_{ij}(k)$ we get the difference
$\gamma_{ij}(k) = h_{ij}(k) - c_{ij}(k)$ which is the fourier
transform of a continuous function in real space. So it is safe to
transform back in real space [to get $\gamma_{ij}(r)$]. Finally, the
dcf's are obtained from the differences $h_{ij}(r) - \gamma_{ij}(r)$. 

While for a system of non-additive hard sphere in three dimensions a
closed form solution to the PY approximation is still lacking, Ahn and
Lebowitz have found an analytic solution of this  approximation for
the WR model (in one and three dimensions). 

The PY approximation consists
of the assumption that $c_{ij}(r)$ does not extend beyond the range of
the potential
\bq
c_{ij}(r)=0~~~~\mbox{for $r>R_{ij}$}~~.
\eq
Combining this with the exact relation (\ref{g12rlalpha})
and using the Ornstein-Zernike equation we are left with
a set of equations for $c_{ij}(r)$ and $g_{ij}(r)$ which have been
solved analytically by Ahn and Lebowitz.

Their solution is parameterized by a parameter $z_0$. They introduce
the following two functions of $z_0$ (which can be written in terms of
elliptic integrals of the first and third kind)
\bq
I_1&\equiv&\int_{z_0}^\infty\frac{dz}{z\sqrt{z^3+4z/z_0-4}}~~,\\
I_2&\equiv&\int_{z_0}^\infty\frac{dz}{\sqrt{z^3+4z/z_0-4}}~~,
\eq 
and define $z_0$ in terms of the partial densities $\rho_1$ and
$\rho_2$ as follows
\bq
\eta\equiv 2\pi\sqrt{\rho_1\rho_2}=\frac{(I_2/2)^3}{\cos I_1}~~.
\eq
They then define the following functions (note that in the last
equality of equation (3.76) in \cite{Ahn74} there is a misprint)
\bq \nonumber 
\bar{c}_{12}(k)&\equiv&-\frac{2}{\sqrt{\rho_1\rho_2}}\sqrt{
\frac{1+Y}{z_0^3Y^3+4Y+4}}\\
&&\times\sin\left[\frac{1}{2}\sqrt{z_0^3Y^3+4Y+4}
\int_{1}^\infty\frac{dz}{(z+Y)\sqrt{z_0^3z^3+4z-4}}\right]~~,\\
\bar{h}_{12}(k)&\equiv&\bar{c}_{12}(k)[1-\rho_1\rho_2\bar{c}_{12}(k)]~~,
\eq
where $Y\equiv (2k/I_2)^2$. 

We also realized that some other misprint should be present in the Ahn
and Lebowitz paper since we have found empirically that the PY
solution (with $k$ in units of $\alpha$) should be  given by 
\bq
c_{12}(k)=\bar{c}_{12}(ks)~~,
\eq
where $s$ is a scale parameter to be determined as follows
\bq
s=-[\bar{h}_{12}(r=0)]^{1/3}
\eq
Notice that for the symmetric case $\rho_1=\rho_2=\rho/2$ and
$\eta=\pi\rho= 0.90316\ldots$ we find $z_0=1$ and $s=1$.  

In Figs. \ref{fig:fig_28748}, \ref{fig:fig_4}, and \ref{fig:fig_65} we
show three cases corresponding to the extreme and one intermediate density.
In the figures, we 
compare the MC simulation data with the PY solution for the partial pair
distribution functions and the partial direct correlation functions.
Our results for the partial pair distribution functions at
$\rho\alpha^3=0.65$ are in good agreement with the ones of Shew and
Yethiraj \cite{Shew96}. The figures show how the like correlation
functions obtained from the PY approximation are the ones that differ
most from the MC simulation data. The difference is more marked in a
neighborhood of $r=0$ and becomes more pronounced as the density increases.

\section{Fit of the data}

From the simulations we found that $c^{(MC)}_{12}(r)<8\times 10^{-3}$
for $r>\alpha$ at all the densities studied. This allows us to say that
$\Delta^c_{12}(r)\simeq 0$ for $r>\alpha$. Moreover we found that both
$\Delta^c_{12}(r)$ for $r<\alpha$, and $\Delta^c_{11}(r)$ are very well 
fitted by Gaussians
\bq \label{gaussianfit11}
\Delta^c_{11}(r)&\simeq&b_{11}\exp[-a_{11}(r+d_{11})^2]~~~~\mbox{for
all $r>0$},\\ 
\label{gaussianfit12}
\Delta^c_{12}(r)&\simeq&b_{12}\exp[-a_{12}r^2]~~~~\mbox{for $0<r<\alpha$},
\eq
In
Figs. \ref{fig:fig_fit12} and \ref{fig:fig_fit11} we show the
behaviors of the parameters of the best fit (\ref{gaussianfit11}) and
(\ref{gaussianfit12}), with density. In order to check the quality of
the fit, we did not use the data at $\bar{\rho}=0.45$ in the
determination of the parameters. The points for $a_{12}$ and
$b_{12}$ are well fitted by a straight line or a parabola. 
As shown in Fig. \ref{fig:fig_fit12} the best parabolae are  
\bq \label{fita12}
a_{12}(\bar{\rho})&=& 0.839+0.096\bar{\rho}-1.287\bar{\rho}^2~~,\\ 
\label{fitb12}
b_{12}(\bar{\rho})&=&-0.155+0.759\bar{\rho}-0.159\bar{\rho}^2~~.
\eq
Fig. \ref{fig:fig_fit11} shows how the parameters for $\Delta^c_{11}(r)$
are much more scattered and hard to fit. The quartic polynomial going
through the five points, for each coefficient, are
\bq \label{fita11}
a_{11}(\bar{\rho})&=&-55.25+504.8\bar{\rho}-1659.\bar{\rho}^2+
2364.\bar{\rho}^3-1236.\bar{\rho}^4~~,\\
b_{11}(\bar{\rho})&=&171.4-1556.\bar{\rho}+5166.\bar{\rho}^2-
7421.\bar{\rho}^3+3906.\bar{\rho}^4~~,\\ \label{fitc11}
d_{11}(\bar{\rho})&=&128.9-1144.\bar{\rho}+3747.\bar{\rho}^2-
5328.\bar{\rho}^3+2782.\bar{\rho}^4~~,
\eq
The difficulty in finding a good fit for these parameters may be
twofold: first we are fitting $\Delta^c_{11}(r)$ with a three (instead
of two) parameters curve and second the partial pair distribution
functions obtained from the Monte Carlo simulation are less accurate
in a neighborhood of the origin (due to the reduced statistics
there). This inaccuracy is amplified in the
process of finding the partial direct correlation functions. Such
inaccuracy will not affect significantly $\Delta^c_{12}(r)$ which has
a derivative very close or equal to zero near the origin, but it
will significantly affect $\Delta^c_{11}(r)$ which is very steep near
the origin.

In order to estimate the quality of the fit we have used the 
simulation data  at $\bar{\rho}=0.45$. From Fig. \ref{fig:fig_fit12}
we can see how the parabolic fit is a very good one. In
Fig. \ref{fig:fig_fit11} the point at $\bar{\rho}=0.45$ gives an
indication of the accuracy of the quartic fit. We have also compared
the pair distribution and direct correlation functions obtained from
the fit with  those from MC: both the like and unlike distribution
functions are well reproduced while there is a visible discrepancy in
the dcf from the origin up to $r=0.5\alpha$.
However we expect that moving on the high density or low
density regions (where the quartic polynomial becomes more steep) the
quality of the fit will get worst. In particular the predicted
negative values for $a_{11}$, in those regions, are completely
unphysical and the fit should not be used to extrapolate beyond the
range $0.28 < \bar{\rho} < 0.65$.

\section{From WR to non additive hard spheres}

In order to see how the structure, and in particular the dcf's  
of the Widom-Rowlinson model change
as one switches on the spheres diameters we have made two additional
Monte Carlo simulations. In the first one we chose
$\rho_1=\rho_2=0.65/R_{12}^3$ and $R_{11}=R_{22}=R_{12}/2$. The
resulting partial pair distribution functions and partial direct
correlation functions are shown in Fig. \ref{fig:fig_65s}. 
From a comparison with Fig. \ref{fig:fig_65} we see how in this case
the switching on of the like diameters causes both $c_{12}(r)$ for
$r<R_{12}$ and $g_{12}(r)$ for $r>R_{12}$ to approach $r=R_{12}$ with
a slope close to zero. 

In the second simulation we chose $\rho_1=\rho_2=0.65/R_{12}^3$ and
$R_{11}=0$, $R_{22}=R_{12}$. The resulting partial pair distribution
functions and partial direct correlation functions are shown in
Fig. \ref{fig:fig_65a}. From a comparison with Fig. \ref{fig:fig_65}
we see how in this case the switching on of the like diameters causes
both $g_{11}(0)$ and $c_{11}(0)$ to increase, and $c_{12}(r)$ to lose
the nearly zero slope at $r=0$. As in the previous case $g_{12}(r)$
for $r>R_{12}$ approaches $r=R_{12}$ with a slope close to zero. The
like $22$ correlation functions for $r>R_{12}$ vary over a range
comparable to the one over which vary the like $11$ correlation
functions of the WR model.

For both these cases there is no analytic solution of the PY
approximation available and a better understanding of the behavior of
the direct correlation functions may help in finding approximate
expressions \cite{GP}.

\section{Conclusions}

In this paper we have evaluated the direct correlation functions of a
Widom-Rowlinson mixture at different densities through  
Monte Carlo simulation and we have studied the possibility of fitting
the difference between MC data and the PY dcf's.  
We found a very good parameterization of
$c_{12}(r)$ for $r<\alpha$ [see equations (\ref{gaussianfit12}) and
(\ref{fita12})-(\ref{fitb12})] and a poorer one for $c_{11}(r)$ [see
equations (\ref{gaussianfit11}) and (\ref{fita11})-(\ref{fitc11})].
The difficulty  in this last case probably arises  from the necessity of
using three parameters [instead of just two needed for parameterizing
$c_{12}(r)$], although it cannot be completely excluded some 
effect of the decreasing precision  of the simulation data near the
origin.

In the last part of the paper we have illustrated with additional 
Monte Carlo data the changes induced in the WR dcf's by a finite size
of the excluded volume of like correlations.
These results are meant to provide a guide in the search of  a 
manageable, simple  
analytical parameterization of  the structure of 
mixtures of non additive hard spheres which is still
not available although highly desirable. 


\bibliography{mcwr}

\begin{figure}[hbt]
\begin{center}
\includegraphics[width=10cm]{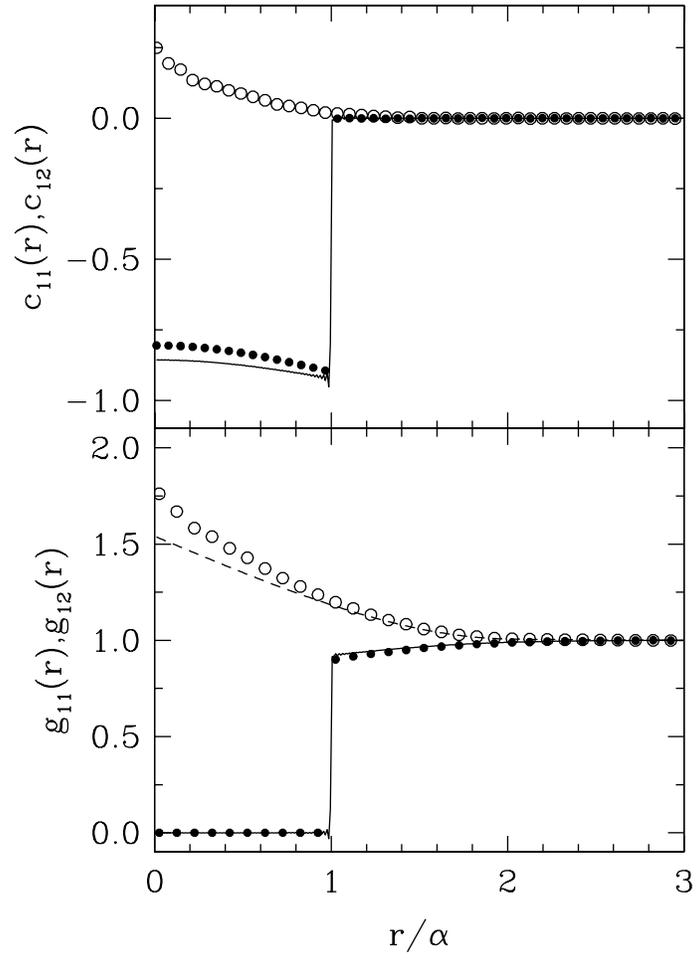}
\end{center}
\caption[]{R. Fantoni and G. Pastore 
\label{fig:fig_28748}
}
\end{figure}

\begin{figure}[hbt]
\begin{center}
\includegraphics[width=10cm]{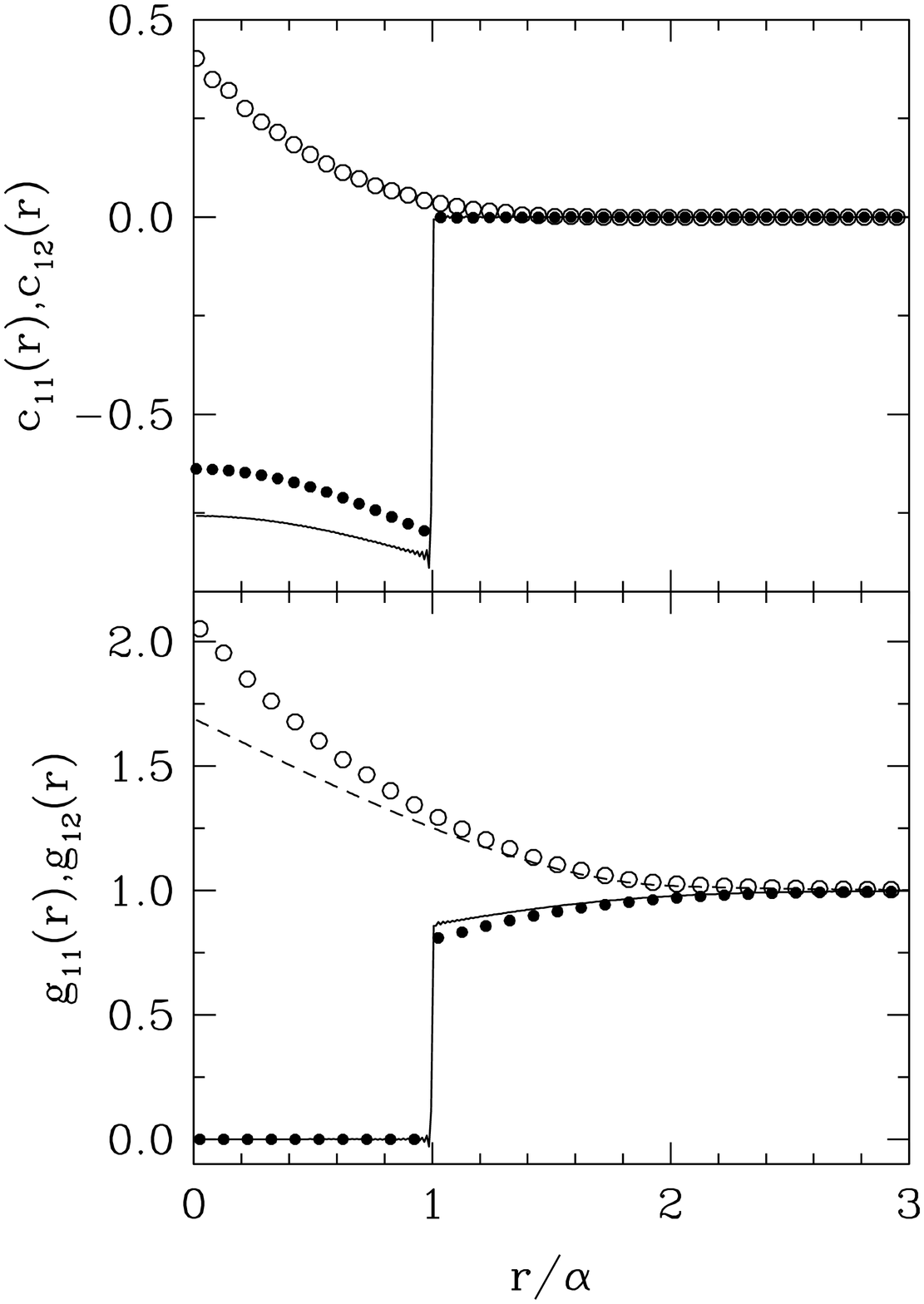}
\end{center}
\caption[]{R. Fantoni and G. Pastore 
\label{fig:fig_4}
}
\end{figure}

\begin{figure}[hbt]
\begin{center}
\includegraphics[width=10cm]{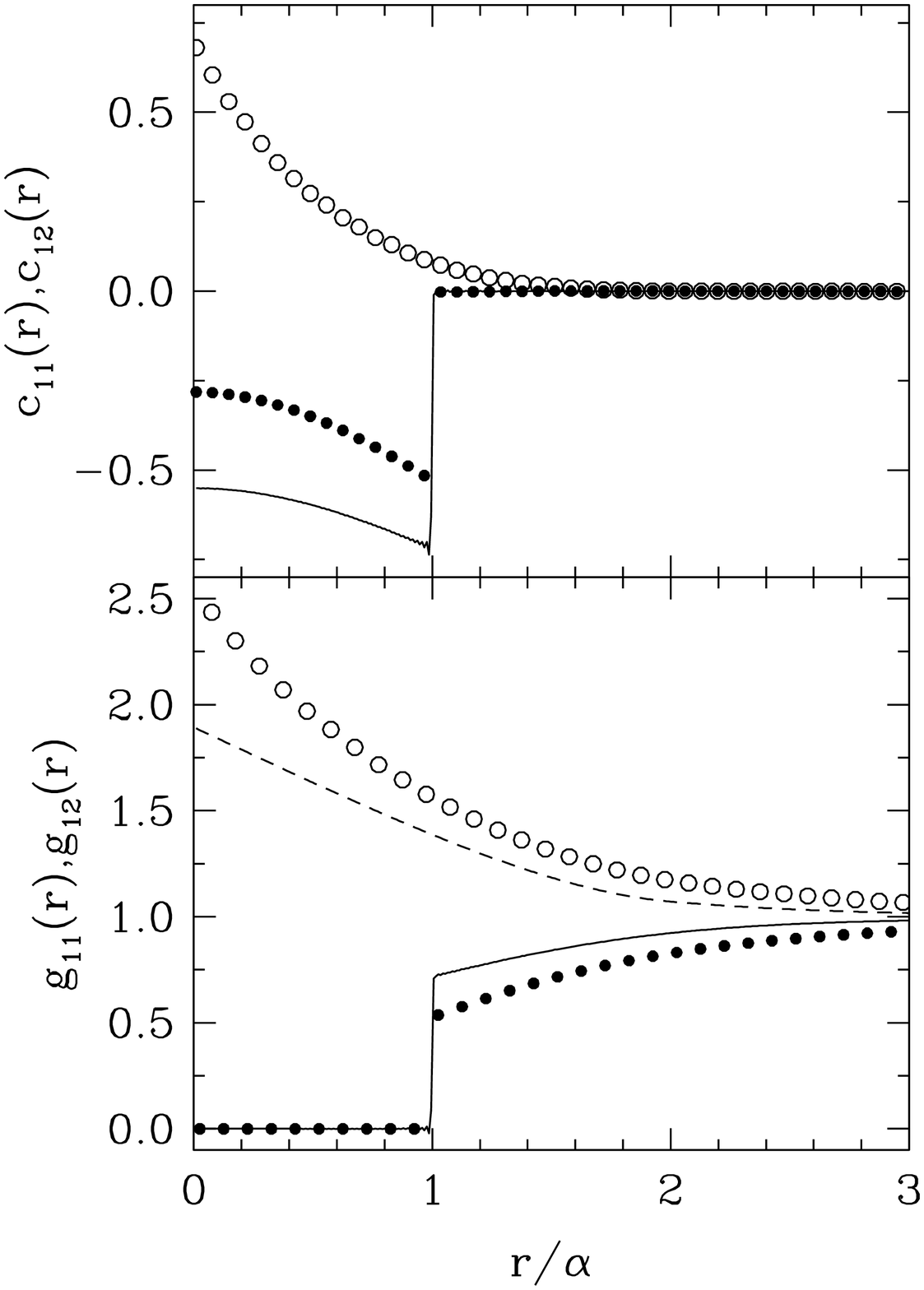}
\end{center}
\caption[]{R. Fantoni and G. Pastore 
\label{fig:fig_65}
}
\end{figure}

\begin{figure}[hbt]
\begin{center}
\includegraphics[width=10cm]{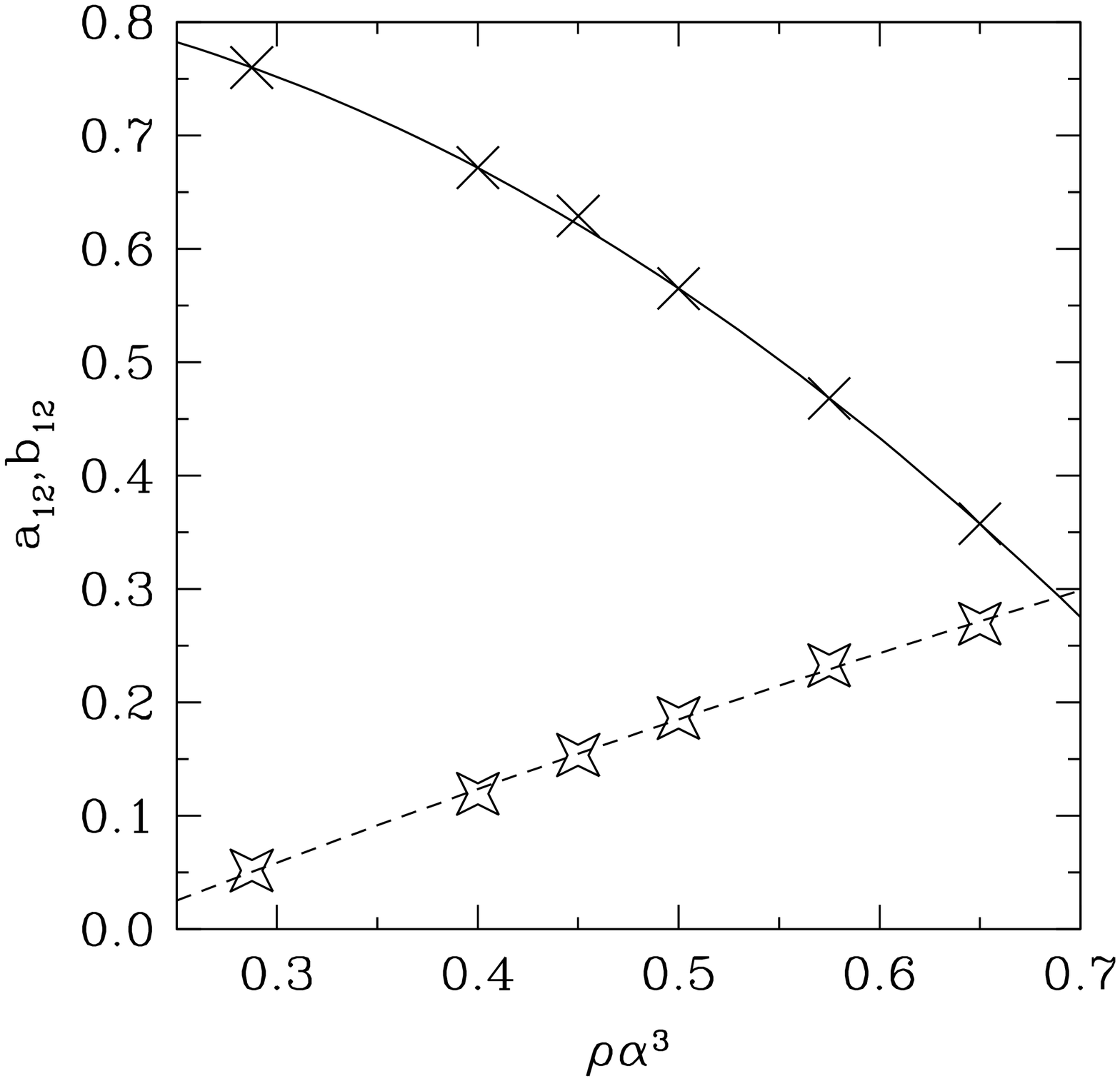}
\end{center}
\caption[]{R. Fantoni and G. Pastore 
\label{fig:fig_fit12}
}
\end{figure}

\begin{figure}[hbt]
\begin{center}
\includegraphics[width=10cm]{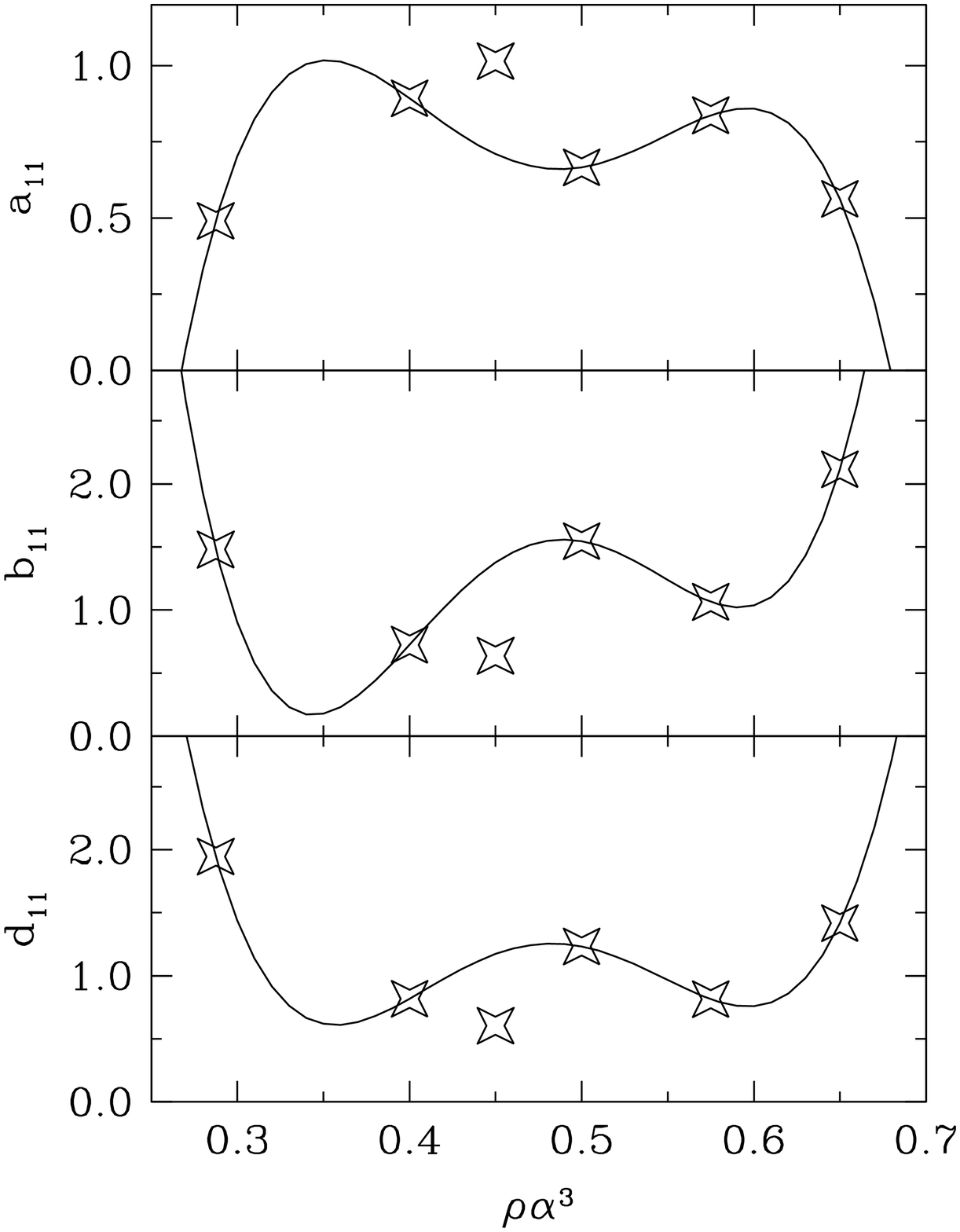}
\end{center}
\caption[]{R. Fantoni and G. Pastore 
\label{fig:fig_fit11}
}
\end{figure}

\begin{figure}[hbt]
\begin{center}
\includegraphics[width=10cm]{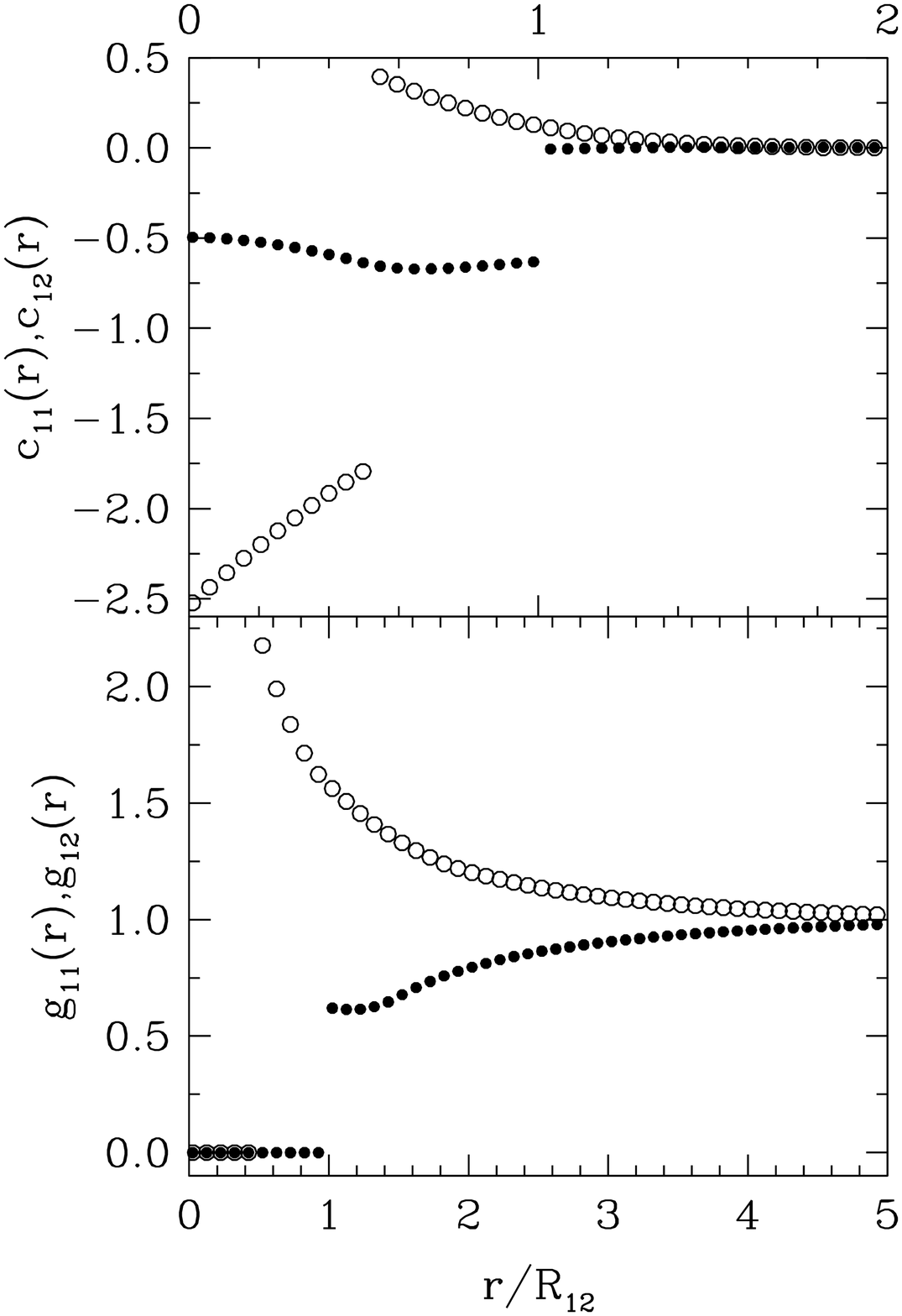}
\end{center}
\caption[]{R. Fantoni and G. Pastore 
\label{fig:fig_65s}
}
\end{figure}

\begin{figure}[hbt]
\begin{center}
\includegraphics[width=10cm]{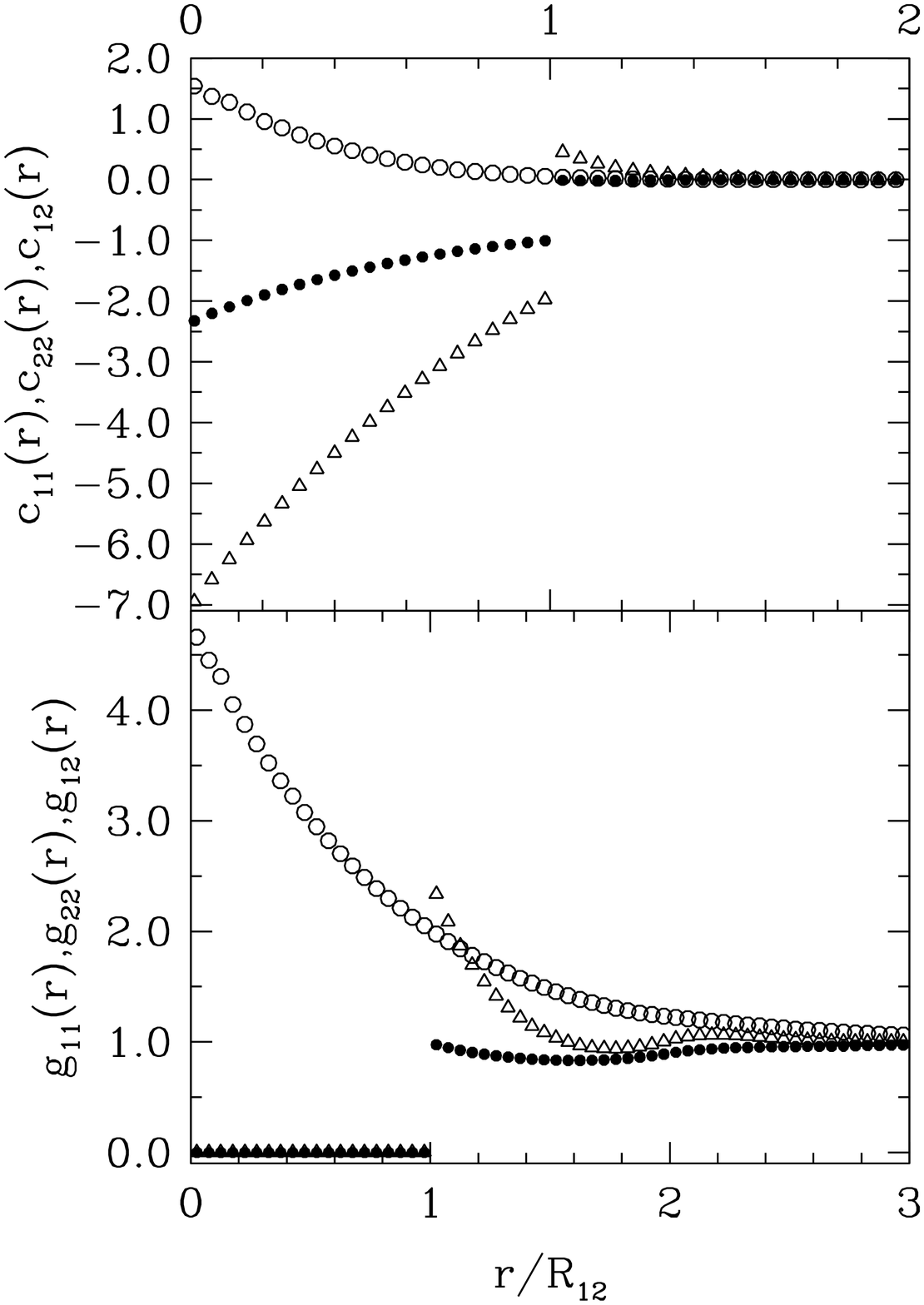}
\end{center}
\caption[]{R. Fantoni and G. Pastore 
\label{fig:fig_65a}
}
\end{figure}

\newpage
\centerline{\bf LIST OF FIGURES}
\begin{itemize}
\item[Fig. \ref{fig:fig_28748}] Top panel: partial direct
correlation functions obtained from the Monte Carlo simulation (points)
with the $c_{12}^{(PY)}(r)$ obtained from the PY approximation (line) 
at a density $\rho\alpha^3=0.28748$. 
Bottom panel:  partial pair distribution
functions obtained from the Monte Carlo simulation compared with the ones
obtained from the PY approximation at the same density. The open
circles and the dashed line:
the like correlation functions. 
Closed circles and the continuous line:   the unlike correlation
functions.

\item[Fig. \ref{fig:fig_4}] Same as in Fig.1 
at a density
$\rho\alpha^3=0.4$.

\item[Fig. \ref{fig:fig_65}] Same as in Fig.1
at a density
$\rho\alpha^3=0.65$.

\item[Fig. \ref{fig:fig_fit12}] We plot, for five different values of
the density, the parameters $a_{12}$ (diagonal crosses) and $b_{12}$ 
(starred crosses) of the best Gaussian fit (\ref{gaussianfit12}) to
$\Delta^c_{12}(r)$ for $r<\alpha$, and fit them with parabolae (lines). 
The parameters at $\rho\alpha^3=0.45$ where not
used for the parabolic fit and give an indication of the
quality of the fit.

\item[Fig. \ref{fig:fig_fit11}] We plot, for five different values of
the density, the parameters $a_{11}, b_{11}$ and $d_{11}$ (stars) of
the best Gaussian fit (\ref{gaussianfit11}) to $\Delta^c_{11}(r)$, and
draw the quartic polynomial (lines) through them. The parameters at
$\rho\alpha^3=0.45$ where not used to determine the quartic polynomial
and  give an indication of the quality of the fit.

\item[Fig. \ref{fig:fig_65s}] Monte Carlo
results at a density $\rho=\rho_1=\rho_2=0.65/R_{12}^3$ for the
partial direct correlation function (on top) and the partial pair
distribution function (below) of a mixture of non additive hard
spheres with $R_{11}=R_{22}=R_{12}/2$. The open circles denote the
like correlation functions. The closed circles denote the unlike
correlation functions. 

\item[Fig. \ref{fig:fig_65a}]Monte Carlo
results at a density $\rho=\rho_1=\rho_2=0.65/R_{12}^3$ for the
partial direct correlation function (on top) and the partial pair
distribution function (below) of a mixture of non additive hard
spheres with $R_{11}=0$ and $R_{22}=R_{12}$. The open circles denote
the like $11$ correlation functions. The open triangles denote the
like $22$ correlation functions. The closed circles denote the unlike
correlation functions. 
\end{itemize}
\newpage
%
\end{document}